\documentclass[authoryear,preprint,10pt]{elsarticle}

\usepackage{microtype}
\usepackage{amsmath}
\usepackage{aas_macros}
\usepackage{float}
\usepackage{url}
\usepackage{caption}
\usepackage{graphicx}%
\usepackage{multirow}%
\usepackage{amsmath,amssymb,amsfonts}%
\usepackage{amsthm}%
\usepackage{mathrsfs}%
\usepackage[title]{appendix}%
\usepackage{xcolor}%
\usepackage{textcomp}%
\usepackage{manyfoot}%
\usepackage{booktabs}%
\usepackage{algorithm}%
\usepackage{algorithmicx}%
\usepackage{algpseudocode}%
\usepackage{listings}%
\usepackage{xspace}%
\usepackage{siunitx}
\usepackage{array} 
\usepackage[colorlinks=true,
            linkcolor=blue,
            citecolor=blue,
            urlcolor=blue]{hyperref}

\newcommand{\tbb}{T_{\rm bb}\xspace}
\newcommand{\vphot}{V_{\rm phot}\xspace}
\newcommand{\epoch}{t_{\rm gw}\xspace}
\newcommand{\vmax}{V_{\rm max}\xspace}
\newcommand{\rmax}{R_{\rm max}\xspace}
\newcommand{\rphot}{R_{\rm phot}\xspace}

\newcommand{\xspec}{{\sc xspec}\xspace}
\newcommand{\grbjet}{\texttt{grbjet}\xspace}

\newcommand{\thetjet}{\theta_{\rm ej}}
\newcommand{\thetaobs}{\theta_{\rm obs}}

\newcommand{\knname}{AT2017gfo\xspace}

\newcommand{\pcyg}{\mbox{P--Cygni}\xspace}
\newcommand{\ep}{E_{\rm p}\xspace}
\newcommand{\liso}{L_{\rm iso}\xspace}

\journal{Journal of High Energy Astrophysics}

\begin{document}

\begin{frontmatter}

\author[inaf-oas]{Ruben Farinelli\corref{cor1}}

\cortext[cor1]{%
\raggedright
Corresponding author. \newline \emph{E-mail addresses}: 
ruben.farinelli@inaf.it (R. Farinelli), fabrizio.cogato@inaf.it (F. Cogato), 
mattia.bulla@unife.it (M. Bulla), psingh@itp.uni-frankfurt.de (P. Singh), giulia.stratta@inaf.it (G. Stratta), andrea.rossi@inaf.it (A. Rossi), 
eliana.palazzi@inaf.it (E. Palazzi), cristiano.guidorzi@unife.it (C. Guidorzi), 
elisabetta.maiorano@inaf.it (E. Maiorano), lorenzo.amati@inaf.it (L. Amati), 
bing.zhang@unlv.edu (B. Zhang), rezzolla@itp.uni-frankfurt.de (L. Rezzolla), 
filippo.frontera@inaf.it (F. Frontera).
}

\author[unibo,inaf-oas]{Fabrizio Cogato}

\author[unife,infn-fe,inaf-oaa]{Mattia Bulla}

\author[itp,inaf-oas]{Paramvir Singh}

\author[inaf-oas,itp]{Giulia Stratta}

\author[inaf-oas]{Andrea Rossi}

\author[inaf-oas]{Eliana Palazzi}

\author[unife,infn-fe,inaf-oas]{Cristiano Guidorzi}

\author[inaf-oas]{Elisabetta Maiorano}

\author[inaf-oas]{Lorenzo Amati}

\author[hk-inst,hk-dep,nca-unine,dpa-unine]{Bing Zhang}

\author[itp,fias,tri]{Luciano Rezzolla}

\author[unife,inaf-oas]{Filippo Frontera}

\affiliation[inaf-oas]{organization={INAF -- Osservatorio di Astrofisica e Scienza dello Spazio},
         addressline={Via P. Gobetti 101},
         postcode={I-40129},
         state={Bologna},
         country={Italy}}
         
\affiliation[unibo]{organization={Dipartimento di Fisica e Astronomia ``Augusto Righi'' -- Alma Mater Studiorum Università di Bologna},
         addressline={Via P. Gobetti 93/2},
         postcode={I-40129},
         state={Bologna},
         country={Italy}}

\affiliation[unife]{organization={Department of Physics and Earth Science, University of Ferrara},
         addressline={Via Saragat 1},
         postcode={I-44122},
         state={Ferrara},
         country={Italy}}
         
\affiliation[infn-fe]{organization={INFN, Sezione di Ferrara},
         addressline={Via Saragat 1},
         postcode={I-44122},
         state={Ferrara},
         country={Italy}}

\affiliation[inaf-oaa]{organization={INAF -- Osservatorio Astronomico d’Abruzzo},
         addressline={Via Mentore Maggini snc},
         postcode={I-64100},
         state={Teramo},
         country={Italy}}

\affiliation[itp]{organization={Institut für Theoretische Physik, Goethe Universität},
         addressline={Max-von-Laue-Str 1},
         postcode={D-60438},
         state={Frankfurt am Main},
         country={Germany}}

\affiliation[hk-inst]{organization={The Hong Kong Institute for Astronomy and Astrophysics, the University of Hong Kong, Pokfulam},
        city={Hong Kong},
              country={P. R. China}}

\affiliation[hk-dep]{organization={Department of Physics and Astronomy, the University of Hong Kong, Pokfulam},
        city={Hong Kong},
            country={P. R. China}}

\affiliation[nca-unine]{organization={Nevada Center for Astrophysics, University of Nevada},
        city={Las Vegas},
         postcode={89154},
         state={Nevada},
         country={USA}}

\affiliation[dpa-unine]{organization={Department of Physics and Astronomy, University of Nevada},
        city={Las Vegas},
         postcode={89154},
         state={Nevada},
         country={USA}}

\affiliation[fias]{organization={Frankfurt Institute for Advanced Studies},
         addressline={Ruth-Moufang-Str. 1},
         postcode={D-60438},
         state={Frankfurt am Main},
         country={Germany}}
         
\affiliation[tri]{organization={School of Mathematics, Trinity College},
         state={Dublin 2},
         country={Ireland}}

\title{The $\ep - \liso$ correlation: A new diagnostic tool for kilonova transients}

\begin{abstract}
The AT2017gfo kilonova transient remains a unique multi-messenger event thanks to 
its proximity ($z=0.00987$) and the possibility to investigate \mbox{time-resolved} spectra, thus providing evidence of r-process nucleosynthesis. The kilonova signal was extensively studied in the spectral and time domains, providing key insights into the chemical composition and physical properties of the ejecta. Here, we report the discovery of a novel correlation between two fundamental observables: the peak energy of the $E\,F_E$ spectrum, $\ep$, and the isotropic-equivalent luminosity, $\liso$. In particular, we show that up to about 2.5 days after the merger, the \knname spectrum evolves according to: $\text{log}_{10} [\ep/\text{eV}] =  -0.13^{+0.02}_{-0.02} + 0.62^{+0.02}_{-0.02} ~\text{log}_{10} [\liso/(10^{41}~\text{erg}~\text{s}^{-1})]~(68\%~{\rm C.L.})$ while in the subsequent epochs, $\ep$ remains almost constant with $\liso$, flattening around $1~\text{eV}$. 
Exploiting simulations from a state-of-the-art radiative transfer code, we demonstrate that our kilonova model inherently predicts this peculiar correlation, hence suggesting a new diagnostic tool for comparing observables against simulations. Future kilonova observations will provide additional insight into the physics behind the $\ep-\liso$ correlation.
\end{abstract}


\end{frontmatter}

\section{An expanding photosphere powering the \knname kilonova}
The detection of the binary neutron star (BNS) merger GW170817 with the LIGO/Virgo interferometer network \cite{Abbott2017a, Abbott2017b}, together with the subsequent observations of a weak short gamma-ray burst (GRB), named GRB170817A \cite{goldstein2017fermi, savchenko2017int} and a bright optical/near-infrared (NIR) kilonova transient, named \knname \cite{coulter2017opt}, has been a landmark event in modern astrophysics, providing crucial insights about compact objects mergers, unattainable from a separate electromagnetic or gravitational wave perspective. In particular, it has confirmed the long-standing conjecture that BNS mergers are behind the phenomenology observed in short GRBs \cite{Eichler89, Narayan92, Rezzolla:2011}.
Furthermore, the temporal evolution of the spectral properties of \knname has provided undisputed evidence of heavy element production, establishing BNS as optimal sites for the nucleosynthesis of elements of the r-process \cite{pian2017nat, smartt2017nat, Kasen2017Natur, watson2019, sneppen2024_hour}.
Within roughly ten days after the BNS merger, the \knname optical/NIR spectra are well described by a cooling blackbody emitted from a homologously expanding photosphere \cite{pian2017nat, watson2019, Waxman2018bb}.

In this work, we focus on the evolution of \knname spectra observed during the first 7.40 days after the BNS merger \citep{pian2017nat, smartt2017nat,  shapee2017magellan, andreoni2017anu, buckley2018salt, nicholl2017soar, chornock2017gemini} with a suite of different instruments, see Table \ref{tab:data}. 
In practice,  following a standard procedure \citep{watson2019, sneppen2024_hour, sneppen2023_nature, Domoto2021ApJ...913...26D, Domoto2022ApJ...939....8D, Gillanders2022MNRAS.515..631G, Perego2022ApJ...925...22P,sneppen0.76}, we adopt a spherically symmetric photosphere powered by the radioactive decay of r-process elements, together with a chemically rich surrounding environment that produces multiple features in the observed blackbody spectra. 
\begin{table}[t]
\centering
\caption{Spectroscopic data set and models of the \knname kilonova transient.}
\resizebox{\textwidth}{!}{
\begin{tabular}{lllll}
\hline
\addlinespace[0.5em]
$\epoch$ & Telescope/Instrument & Spectral Range & Reference & Spectral Model
\\ 
(days)& &(nm)& \\ 
\addlinespace[0.5em]
\hline
\addlinespace[0.5em]
0.49 & Magellan/LDSS & 360--700   & \cite{shapee2017magellan} & \grbjet \\
0.53 & Magellan/MagE & 370--1000  & \cite{shapee2017magellan} & \grbjet\\
0.92 & ANU/WiFeS     & 200--900   & \cite{andreoni2017anu} & \grbjet\\
1.17 & SALT/RSS      & 360--810   & \cite{buckley2018salt} & \grbjet + \pcyg \\
1.43 & VLT/X-Shooter & 330--2250  & \cite{pian2017nat,smartt2017nat} & \grbjet + \pcyg  + NIR \\
1.47 & SOAR+Gemini & 390--1800  & \cite{nicholl2017soar,chornock2017gemini} & \grbjet + \pcyg + NIR  \\
2.40 & VLT/X-Shooter & 330--2250  & \cite{pian2017nat,smartt2017nat} & \grbjet + \pcyg + NIR \\
3.40 & VLT/X-Shooter & 330--2250  & \cite{pian2017nat,smartt2017nat} & \grbjet + \pcyg + NIR \\
4.40 & VLT/X-Shooter & 330--2250  & \cite{pian2017nat,smartt2017nat} & \grbjet + \pcyg + NIR \\
5.40 & VLT/X-Shooter & 330--2250  & \cite{pian2017nat,smartt2017nat} & \grbjet + \pcyg + NIR \\
6.40 & VLT/X-Shooter & 330--2250  & \cite{pian2017nat,smartt2017nat} & \grbjet + \pcyg + NIR \\
7.40 & VLT/X-Shooter & 330--2250  & \cite{pian2017nat,smartt2017nat} & \grbjet + \pcyg + NIR \\
\addlinespace[0.5em]
\hline
\end{tabular}
}
\label{tab:data}
\end{table}

To model the continuum of \knname we exploit the \grbjet model\,\footnote{Originally developed for GRB jet emission and freely available in the \xspec \citep{arnaud1996} package.}\citep{farinelli2021}, assuming a blackbody emission from a sphere ($\theta_{\rm ej} = 90^{\circ}$ and $\theta_{\rm obs} = 0^{\circ}$, see Table \ref{tab:model}) of radius $\rphot$ expanding with a velocity $\vphot$ and radiating at a certain comoving frame temperature $\tbb$. 
The blackbody emission is thus modelled in the comoving frame of the expanding photosphere, thereby including the correction factors arising from the subrelativistic expansion of the kilonova ejecta \cite{sneppen23_bb}.
The luminosity distance of the emitting source is determined assuming a Hubble constant value of $70~\rm{km}^{-1} {\rm s}^{-1} \rm{Mpc}^{-1}$ and considering the \knname-host galaxy redshift $z=0.00987$. 
Under the assumption of homologous expansion, the velocity of any given part of the ejecta is proportional to its distance from the centre \cite{Eastman96, Branch00, Dessart05} so that the radius of the photosphere can be directly inferred from the time after the BNS merger $\epoch$, namely $\rphot \equiv \vphot ~ \epoch$.

Detailed analysis of the \knname spectral properties \citep{sneppen2024_hour} has revealed the emergence of various spectral features in different wavelength regimes as a function of the evolutionary state of the kilonova ejecta. 
Therefore, together with the photospheric component, the kilonova ejecta comprises a resonant scattering region which extends above the photosphere up to a certain radius $\rmax$ ($\equiv \vmax ~ \epoch$) and is responsible for the formation of several \pcyg profiles \citep{watson2019, sneppen2023_nature, jeffrey1990, kasen02}.

\begin{table}[t]
\caption{Model components and parameters used to fit the spectra of \knname. Parameters between square brackets have been kept frozen during the fit. See Table \ref{tab:data} for the use of single or multi-component at each observing epoch.}
\centering
\resizebox{\textwidth}{!}{
\begin{tabular}{lll}
\hline
\addlinespace[0.5em]
Parameter & Description & Fit setup  \\ 
\addlinespace[0.5em]
\hline
\addlinespace[0.5em]
\multicolumn{3}{l}{Expanding ejecta (\grbjet)}\\
\addlinespace[0.5em]
\hline
\addlinespace[0.5em]
\rule{0pt}{10pt}
$z$ & Host galaxy redshift & [0.00987] \\
\ $\thetaobs$ & Observer viewing angle (deg) & [0] \\
\ $\thetjet$ & Ejecta half opening angle (deg) & [90]  \\
\ $\Gamma$ & Photosphere Lorentz factor & $\equiv (1-V^2_{\rm phot})^{-1/2}$ \\
\ $\rphot$ & Photospheric radius (cm) & $\equiv \vphot \epoch$  \\
\ $\tbb$ & Comoving blackbody temperature (K) & free  \\
\addlinespace[0.5em]
\hline
\addlinespace[0.5em]
\multicolumn{3}{l}{\pcyg profile}\\
\addlinespace[0.5em]
\hline
\addlinespace[0.5em]
\rule{0pt}{10pt}
$t$ & Time since GW trigger (s) & $[\epoch]$ \\
\ $\vphot$ &  Ejecta velocity at photospheric radius ($c$) & free \\
\ $\vmax$ & Ejecta velocity at outer radius ($c$)  & free \\
\ $\tau$ & Optical depth of resonant region &  free\\
\ $\xi$ & Occultation parameter &  free \\
\addlinespace[0.5em]
\hline
\addlinespace[0.5em]
\multicolumn{3}{l}{NIR Gaussian emission line ($\times$ 2)} \\
\addlinespace[0.5em]
\hline
\addlinespace[0.5em]
\rule{0pt}{10pt}
$E_{\lambda}$ & Line centroid ($\si{\angstrom}$) & free \\
\ $\sigma_{\lambda}$ & Line width ($\si{\angstrom}$) &  free \\
\ $N_{\lambda}$ & Line normalization (erg~cm$^{-2}~$s$^{-1}$~$\si{\angstrom}^{-1}$) & free \\
\addlinespace[0.5em]
\hline
\end{tabular}
}
\label{tab:model}
\end{table}

In particular, we profit from freely available packages\,\footnote{Original version: \url{https://github.com/unoebauer/public-astro-tools}}\,\footnote{Updated version: \url{https://github.com/Sneppen/Kilonova-analysis}} to model three \pcyg profiles arising from the Sr$^{+}$ triplet around 1 $\mu{\rm m}$.
Within the spherically-symmetric framework, the Sr$^{+}$ triplet at 1.004 $\mu$m, 1.033 $\mu$m and 1.091 $\mu$m is modeled with an optical depth $\tau$ of the resonant scattering region, and weighted by the coefficients $1/13.8$, $8.1/13.8$ and $4.7/13.8$, respectively, as obtained from TARDIS simulations \citep{watson2019}.
Moreover, to account for the asymmetry of the \pcyg profile arising from processes which are not taken into account by the standard line-formation model, we include a phenomenological parameter $\xi$ which enhances the emission part of the \pcyg profile \cite{sneppen2023_h0}.
Finally, we consider two NIR Gaussian emission lines around 1.55 $\mu$m and 2.05 $\mu$m, whose origin is still unclear \citep{watson2019}. 
We do not include other \pcyg profiles, such as the YII feature around $0.76\ \mu$m \citep{sneppen0.76}, the SrII feature around $0.41\ \mu m$ \citep{watson2019}, or the NIR features around $1.4\ \mu $m, $1.6\ \mu$m and $2.1\ \mu$m \citep{Domoto2022ApJ...939....8D, Hotokezaka2023, sneppen2024_hour}.
Table \ref{tab:model} outlines the model parameters and the configuration used in the fitting process.

We analyse the temporal evolution of the \knname spectra throughout 12 different epochs\,\footnote{Available at \url{https://github.com/Sneppen/Kilonova-analysis}}, ranging from 0.49 to 7.40 days after the BNS merger, which were obtained thanks to a global network of ground-based astronomical observatories (see Table \ref{tab:data}).
To fit such a complex spectroscopic dataset within a Markov Chain Monte Carlo framework \citep{emcee}, we implement a custom software able to access the object-shared library from the source code of the \grbjet model and construct the likelihood function including the \pcyg and the NIR Gaussian profiles.
The posterior probability is sampled with 100 walkers and $N>5000$ steps, applying a uniform prior on each set of model parameters. With a burn-in of 2000 steps, the best-fit values are obtained by taking the median and the 68$\%$ confidence level of the marginalised posterior distributions.

Our analysis aims to probe the properties of the expanding photosphere without contamination from narrow-band spectral features; hence, we focus on the temporal evolution of the best-fit blackbody continuum. In particular, the evolution of the kilonova ejecta is tracked by measuring the velocity of the expanding photosphere $\vphot$ and the comoving frame temperature $\tbb$ of the associated blackbody, as reported in Table \ref{tab:bf_rvt_homologous}.
At $\epoch<0.6$\,days, the available observations \citep{shapee2017magellan} do not offer sufficient spectral coverage to include the blackbody peak of \knname ($\tbb > 8000\,{\rm K}$, $\lambda_{\rm peak} < 360\,{\rm nm}$), and no \pcyg profiles have been detected so far.
Around $\epoch \sim 1$ days \citep{andreoni2017anu, buckley2018salt}, the blackbody temperature decreases to $\tbb \sim 5500$\,K, thus positioning the spectral peak ($\lambda_{\rm peak} \sim 530\,{\rm nm}$) within the spectral coverage of the available observations. Although the spectrum remains featureless up to 0.92 days after the BNS merger, at $\epoch = 1.17$\,days the \pcyg profile at $1 \mu$m has emerged for the first time \citep{sneppen2024_hour}.
From $\epoch \sim 1.4$ days onward \citep{pian2017nat,smartt2017nat,nicholl2017soar,chornock2017gemini}, the blackbody spectrum largely falls into the spectral coverage of the \knname dataset. All spectra are analysed with the \pcyg profile at $1~\mu$m and two Gaussian emission lines around $1.55~\mu$m and $2.05~\mu $m.
The first VLT/X-Shooter and the SOAR+Gemini observations are almost contemporaneous at $\epoch \sim 1.45$ days, with an average blackbody temperature $\tbb \sim 4600$ K. 

Ignoring the Magellan/LDSS spectrum at $\epoch = 0.49$\,days which offers the poorest spectral coverage among the available dataset, the subsequent evolution of the \knname spectra reveals that the photospheric velocity $\vphot$ drops by about an order of magnitude between $0.53$\,days and $7.40$\,days after the BNS merger, namely from 0.3\,$c$ to 0.04\,$c$. 
Our findings agree with previous studies of \knname spectra \citep{pian2017nat, smartt2017nat, Kasen2017Natur, sneppen2024_hour, Waxman2018bb}.

Typically, the optical/NIR spectra of astrophysical sources are studied in terms of the spectral flux density $F_{\lambda}$ [erg~cm$^{-2}$~s$^{-1}$~\si{\angstrom}$^{-1}$] as a function of wavelength $\lambda$ [\si{\angstrom}].
Our analysis of \knname is carried out in the $F_{\lambda}$\,-\,$\lambda$ space, fitting the spectra in the comoving frame of the expanding photosphere. However, following the standard approach in X/$\gamma$-ray astronomy, we examine the \knname evolution in units of photon energy $E$ [eV] and radiant flux $E\,F_E$ [erg~cm$^{-2}$~s$^{-1}$], which is equivalent to $\lambda \,F_{\lambda}$.

\begin{figure}[!h]
\centering
\includegraphics[width=\textwidth]{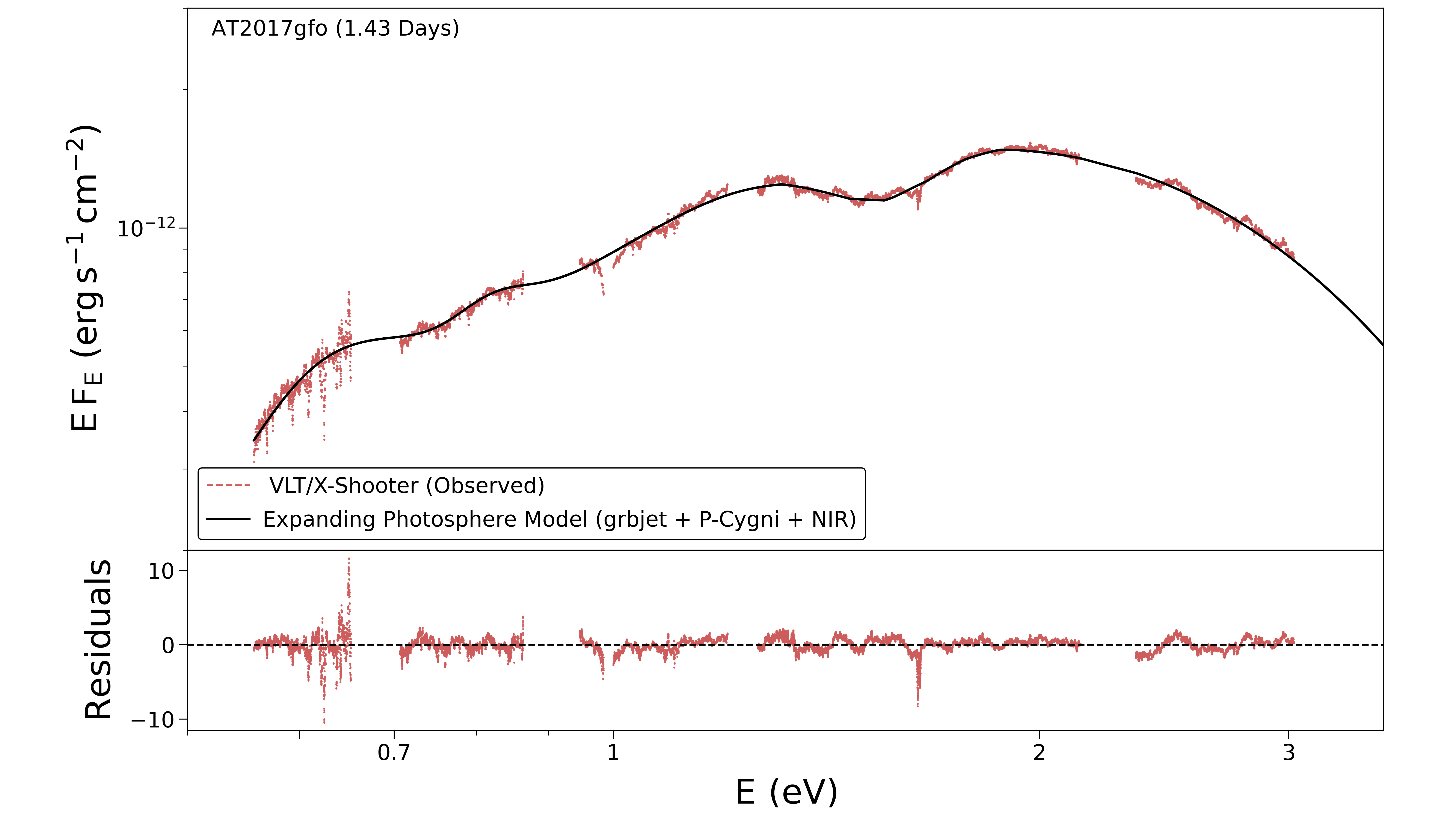}
\caption{Spectrum of VLT/X-Shooter for \knname at epoch $\epoch=1.43$ days in $E\,F_E\,$ units along with the best-fit model and the associated residuals (in units of $\sigma$).}
\label{nuFnu_Fit_Residual}
\end{figure}

An example of the performance of our fitting procedure is shown in Figure \ref{nuFnu_Fit_Residual}, where we present the best-fit model associated with the first VLT/X-Shooter observation at $\epoch \sim 1.43$\,days.
\begin{table}[t]
\caption{Main best-fit parameters obtained from the fit of \knname spectra throughout 12 different epochs after the BNS merger.}
\centering
\resizebox{\textwidth}{!}
{
\begin{tabular}{lllll}
\hline
\addlinespace[0.5em]
$\epoch$ & $\vphot$  & $\vmax$  & $\tbb$ & $\rphot \equiv \vphot ~ \epoch$  \\
(days) & ($c$)  & ($c$) & (K) & $(10^{12}~{\rm cm})$  \\
\addlinespace[0.5em]
\hline
\addlinespace[0.5em]
0.49 & 0.240$\pm$0.001 & --                    & 11129$\pm$30 & 292$\pm$1   \\
0.53 & 0.300$\pm$0.001 & --                    & 8143$\pm$1  & 415$\pm$1   \\
0.92 & 0.270$\pm$0.005  & --                   & 5593$\pm$50 & 635$\pm$1   \\ 
1.17 & 0.230$\pm$0.005  & 0.380$\pm$0.005       & 5479$\pm$50 & 700$\pm$10   \\
1.43  & 0.230$\pm$0.001  & 0.360$\pm$0.001       & 4699$\pm$5  & 842$\pm$2   \\ 
1.47  & 0.220$\pm$0.001  & 0.350$\pm$0.001       & 4659$\pm$6  & 875$\pm$2   \\
2.40  & 0.230$\pm$0.001  & 0.270$\pm$0.002       & 3152$\pm$8 & 1426$\pm$6   \\
3.40  & 0.160$\pm$0.001  & 0.240$\pm$0.002       & 2998$\pm$8 & 1432$\pm$8    \\
4.40  & 0.140$\pm$0.001  & 0.230$\pm$0.002       & 2737$\pm$6  & 1585$\pm$8   \\
5.40  & 0.080$\pm$0.001  & 0.210$\pm$0.004       & 3140$\pm$12 & 1171$\pm$9   \\
6.40  & 0.060$\pm$0.001  & 0.24$\pm$0.04        & 3124$\pm$15 & 987$\pm$12   \\
7.40  & 0.040$\pm$0.001  & 0.24$\pm$0.01        & 3065$\pm$5  & 860$\pm$4   \\
\addlinespace[0.5em]
\hline
\end{tabular}
}
\label{tab:bf_rvt_homologous}
\end{table}

\section{The $\ep-\liso$ correlation as a probe for the blackbody evolution}
The key advantages of the proposed approach are twofold. First, representing the spectrum in these physical units emphasises the photon energy around which the astrophysical source emits most of its power. Second, any evolution in the source spectrum can be readily detected, as shown in the case of X-ray binaries hosting a black hole \citep{montanari2009, farinelli2013}.

\begin{figure}[!h]
\centering
\includegraphics[width=\textwidth]{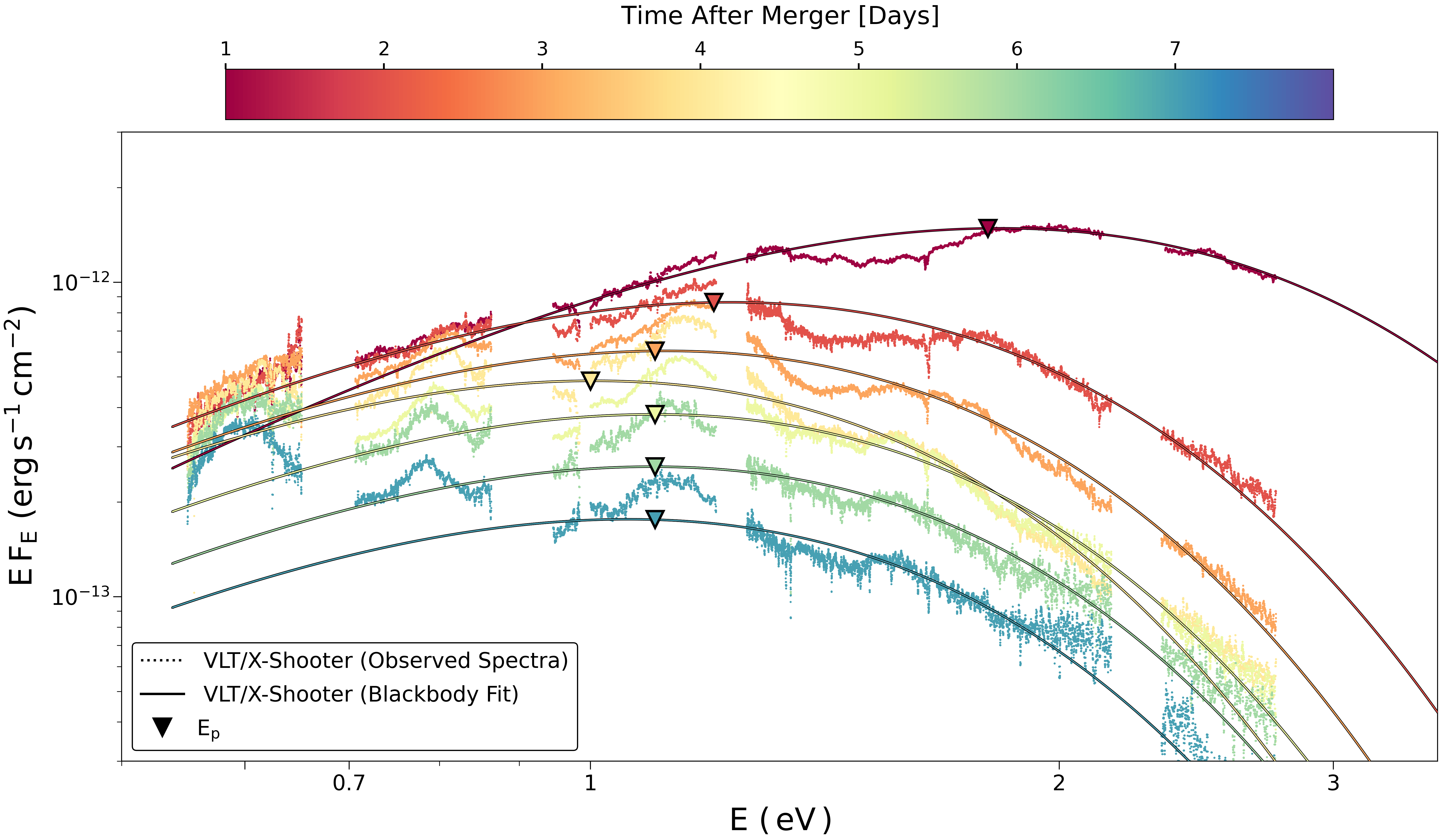}
\caption{Spectra of \knname as observed by VLT/X-Shooter in $EF_E$ units (erg cm$^{-2}$~s$^{-1}$). The epochs not observed with VLT/X-Shooter (see Table 1) are not shown to enhance the visualisation of the temporal evolution of $\ep$.
The subrelativistic blackbody continuum extracted from the \texttt{grbjet} component of the best-fit model is superimposed on the observational data, and the corresponding $\ep$ value is shown with a triangle for each spectral epoch. 
}
\label{fig:nuFnu_XShooter}
\end{figure}

Figure \ref{fig:nuFnu_XShooter} shows the $E\,F_E$ spectra of \knname as observed by the VLT/X-Shooter instrument \citep{pian2017nat, smartt2017nat}, and the temporal evolution of the blackbody component of the best-fit model. Deviations of the observed $E\,F_E$ spectra to the blackbody component are modelled with a \pcyg profile at $\sim1.2$\,eV and two NIR Gaussian features around $0.7$\,eV.

This perspective on the \knname spectral evolution clearly shows that, as expected for a cooling blackbody, a decline in the bolometric flux $S_{\text{bolo}}$ [erg~cm$^{-2}$~s$^{-1}$] of the photosphere is accompanied by a corresponding shift in the peak energy $\ep$ [eV] of its $E\,F_E$ spectrum towards lower energies.
Thus, in the rest frame of the kilonova, from the blackbody best-fit continuum we infer the peak energy $\ep$ [eV] of the $E\,F_E$ spectrum and the isotropic-equivalent luminosity $\liso = S_{\text{bolo}} \times 4\pi D_L^2$ [erg~s$^{-1}$]. 

Because neither $\ep$ nor $\liso$ is a direct fit parameter, we sample the posterior distribution and, for each spectral realisation, infer the $\ep$ and $\liso$ values from the model's blackbody continuum.
This procedure yields a posterior distribution for both $\ep$ and $\liso$, from which we estimate their best-fit values and uncertainties by taking the median and the 68$\%$ confidence level of the corresponding marginalised distributions. The best-fit values of $\ep$ and $\liso$ are mainly dictated by the comoving blackbody temperature $\tbb$ and the photospheric radius $\rphot$, with an additional Doppler factor correction. Therefore, the tight constraints on the spectral model parameters (see Table \ref{tab:bf_rvt_homologous}) reflect in $\sim1\%$ statistical errors on the inferred $\ep$ and $\liso$ value.
\begin{table}[t]
\caption{$\ep$ and $\liso$ best-fit values with associated 1-$\sigma$ statistical errors derived from the subrelativistic blackbody component of the \knname spectra.} 
\centering
\begin{tabular}{lll}
\hline
\addlinespace[0.5em]
$\epoch$ (days) &   $\ep$ (eV) & $\liso$ ($10^{41}$ erg~s$^{-1}$) \\ 
\addlinespace[0.5em]
\hline
\addlinespace[0.5em]
0.49 &  4.35$\pm$0.01 & 17.5$\pm$0.2     \\ 
 0.53 &   3.35$\pm$0.02 & 11.8$\pm$0.8  \\ 
 0.92 &  2.23$\pm$0.02 & 5.4$\pm$0.3  \\ 
  1.17 &  2.14$\pm$0.03 & 5.6$\pm$0.4  \\ 
1.43  &  1.81$\pm$0.01  & 4.45$\pm$0.09 \\ 
 1.47 &  1.836$\pm$0.003  & 4.28$\pm$0.03  \\ 
2.40 &  1.24$\pm$0.01  & 2.37$\pm$0.09  \\ 
3.40 &  1.113$\pm$0.004  & 1.63$\pm$0.03  \\ 
4.40  & 1.005$\pm$0.003  & 1.28$\pm$0.02 \\ 
5.40  &  1.09$\pm$0.01  & 1.03$\pm$0.04  \\ 
6.40  &  1.09$\pm$0.01  & 0.71$\pm$0.04 \\ 
7.40  &  1.07$\pm$0.01  & 0.45$\pm$0.02  \\ 
\addlinespace[0.5em]
\hline
\end{tabular}
\label{tab:bf_epliso}
\end{table}

\begin{figure}[!h]
\centering
\includegraphics[width=\textwidth]{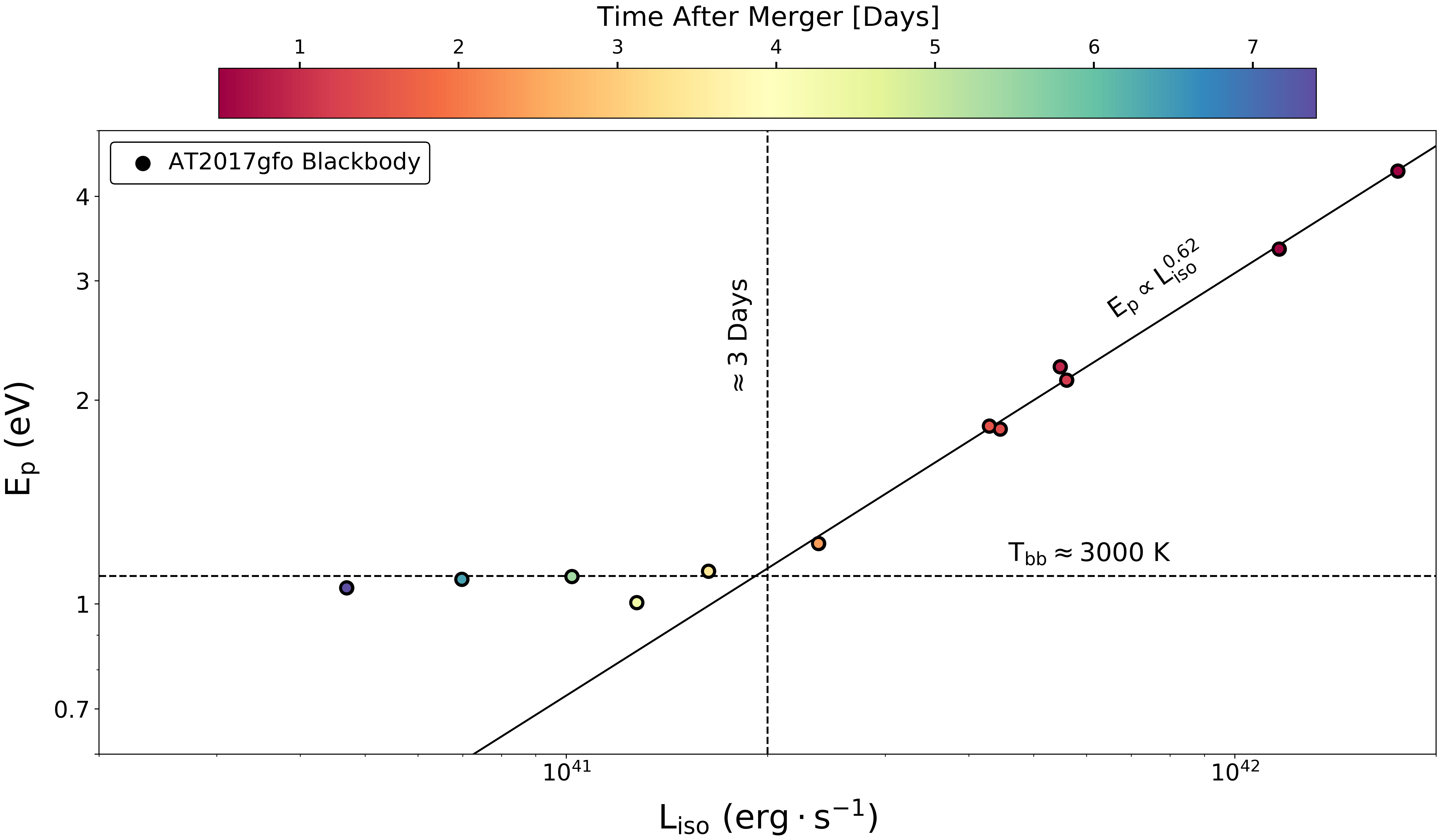}
\caption{The $\ep-\liso$ correlation for the blackbody component in the \knname spectra. Temporal epochs range from 0.49 to 7.40 days after the BNS merger, and the estimated uncertainties on both quantities are of the order of symbol size (see Table \ref{tab:bf_epliso}).}
\label{ep_liso_data}
\end{figure}

Table \ref{tab:bf_epliso} presents the results of this analysis.
Our method probes the evolution of the blackbody photosphere using two ``quasi-empirical" observables.
We refer to $\ep$ and $\liso$ as ``quasi-empirical" observables because, for an ideal blackbody spectrum sampled over a sufficiently broad wavelength range, these values can be measured directly from the observed data. However, in the case of \knname, the values of $\ep$ and $\liso$ cannot be directly extracted from the observed spectra, as the blackbody emission is hidden below a complex combination of emission and absorption features.
Therefore, we rely on the blackbody component of the best-fit model of \knname spectra to trace the evolution of the expanding photosphere in terms of $\ep$ and $\liso$.
In Figure \ref{ep_liso_data}, we observe that \knname follows a peculiar trajectory within the $\ep - \liso$ plane. Up to three days after the merger, encompassing the first seven spectral epochs, \knname follows a tight correlation\footnote{Applying a D'Agostini likelihood \citep{dagostini2005, amati2006}, we infer the extrinsic scatter as $\sigma_{\rm ext} \sim 0.01$.} measured as:
\begin{equation}
    \text{log}_{10} \bigg[\frac{\ep}{\text{eV}}\bigg] \ = \  -0.13^{+0.02}_{-0.02} \ +\ 0.62^{+0.02}_{-0.02} ~\text{log}_{10} \bigg[\frac{\liso}{10^{41} \ \text{erg}~\text{s}^{-1}}\bigg] \, .
\end{equation}
Conversely, during the last five spectral epochs, the correlation settles around $\ep \sim 1$ eV, corresponding to a comoving frame blackbody temperature $\tbb\sim3000$ K.
We interpret this trajectory in the $\ep-\liso$ plane in terms of a physical transition in the photospheric expansion. More specifically, as the photosphere evolves, the expanding material traverses outward, and the blackbody's temperature decreases. Consequently, the material at the boundary of the ejecta becomes progressively optically thin.

At temperatures around 3000 K, the opacity of the outer layers drops significantly, and the observer measures a progressive inward recession of the photospheric boundary, also known as the ``\,receding photosphere\," mechanism \cite{Kasen2013, Tanaka2020, kato24, Yang2024}. 
Under the blackbody assumption, the peak energy $\ep$ serves as a proxy for the comoving frame temperature $\tbb$ of the photosphere.
This explains the emergence of two distinct phases in the \knname kilonova.
Initially, the photosphere expands and radiates like a cooling blackbody.
Around $\tbb \sim 3000$\,K, the boundary of the photosphere becomes optically thin and begins to recede, marking the transition to the second phase.

\section{A new diagnostic tool for kilonova transients}
Since the radiative mechanism of the expanding photosphere is physically linked to the dynamical properties of the ejecta that power the kilonova, the $\ep-\liso$ correlation provides a novel and powerful tool for inferring the physical properties of the ejecta, namely its mass, velocity, and electron fraction. All of these quantities are very hard to quantify and yet provide essential information on the properties of the neutron stars involved in the merger.
In this work, we illustrate how the inferred values of $\ep$ and $\liso$ can be used to effectively test a grid of kilonova models derived from the 3D Monte Carlo radiative transfer code \texttt{POSSIS} \citep{Bulla2019POSSIS, Bulla2023POSSIS}, providing additional information to accurately simulate the dynamical evolution of the ejecta material after a BNS merger.
\begin{table*}[t]
\captionsetup{width=\textwidth}
\caption{\texttt{POSSIS} grid parameter space.}
\centering
\begin{tabular}{ll}
\hline
\addlinespace[0.5em]
Parameters & Grid steps  \\ 
\addlinespace[0.5em]
\hline
\addlinespace[0.5em]
$m_{\rm dyn}$ ($M_\odot$) & $[0.001,0.005,0.010,0.020]$   \\ 
$\bar{v}_{\rm dyn}$ ($c$) & $[0.12,0.15,0.20,0.25]$   \\ 
$\bar{Y}_{e,\,\rm dyn}$ & $ [0.15,0.20,0.25,0.30]$   \\ 
$m_{\rm wind}$ ($M_\odot$) & $[0.01,0.05,0.09,0.13]$  \\ 
$\bar{v}_{\rm wind}$ ($c$) & $[0.03,0.05,0.10,0.15]$ \\ 
$\bar{Y}_{e,\,\rm wind}$ & $[0.20,0.30,0.40]$  \\
\addlinespace[0.5em]
\hline
\end{tabular}
\label{tab:POSSIS_Grid_ParameterSpace}
\end{table*}
In particular, we use the latest version of the code \cite{Bulla2023POSSIS} where key properties of the ejecta like nuclear heating rates \cite{Rosswog2024}, thermalization efficiencies \cite{Barnes2016, Wollaeger2018} and wavelength-dependent opacities \cite{Tanaka2020} are a function of time and local properties of the ejecta (e.g., density, temperature, and chemical composition). Monte Carlo photon packets are tracked along their propagation throughout the ejecta -- assuming its homologous expansion -- and interact with matter via electron scattering and bound-bound transitions (bound-free and free-free transitions are subdominant at the epochs and wavelengths relevant for kilonova emission \cite{Tanaka2020}).

We profit from a grid of 3072 models presented in \cite{Ahumada25} and created from a baseline axially symmetric ejecta model like the one in \cite{Bulla2023POSSIS}.  
Briefly, a first dynamical ejecta component is modelled with the following density profile \citep{Perego2017,Kiuchi2017,Radice2018,Hotokezaka2018,Kawaguchi2020}:
$$\rho(r,\theta)\propto \sin^2\theta\,r^{-\alpha}$$ 
where $\theta$ is the polar angle, $\alpha=4$ in the range $0.1\,c\leq r/t\leq 0.4\,c$, and $\alpha=8$ at larger radii.
Also, the angular profile for the electron fraction is modelled as follows \citep{Radice2018,Setzer2023}:
$$Y_{\rm e}(\theta)\propto \cos^2\theta\, .$$ 
A second disk-wind ejecta component is modelled assuming spherical symmetry, uniform chemical composition, i.e., fixed $Y_{\rm e}$, and the following power law density profile \citep{Kawaguchi2020}:
$$\rho(r)\propto r^{-3} \, .$$
For each ejecta component $i$, the mass $m_i$, the average velocity $\bar{v}_i$, and the average electron fraction $\bar{Y}_{e,i}$, are varied as free parameters within the grid steps summarised in Table \ref{tab:POSSIS_Grid_ParameterSpace}. While spectra from \cite{Ahumada25} are computed for 11 viewing angles from a face-on to an edge-on view of the system, here we select only those relative to an observer at $\theta_{\rm obs} \sim 25^{\circ}$ ($\cos\theta_{\rm obs}=0.9$), i.e., a value compatible with the one inferred for GW170817 from the observed superluminal motion of the jet \citep{Mooley22}. 
These results in $4^5 \times3=3072$ kilonova ejecta models\,\footnote{An earlier version of this grid was presented in \cite{Anand2023} with $\bar{Y}_{e,\,\rm wind} = 0.3$}, providing a total of $36\,864$ kilonova spectra when accounting for the $12$ different observational epochs considered in this work. We refer the reader to \cite{Bulla2023POSSIS,Ahumada25} for a more detailed description of the model. 

\begin{figure}[!h]
\centering
\includegraphics[width=0.82\textwidth]{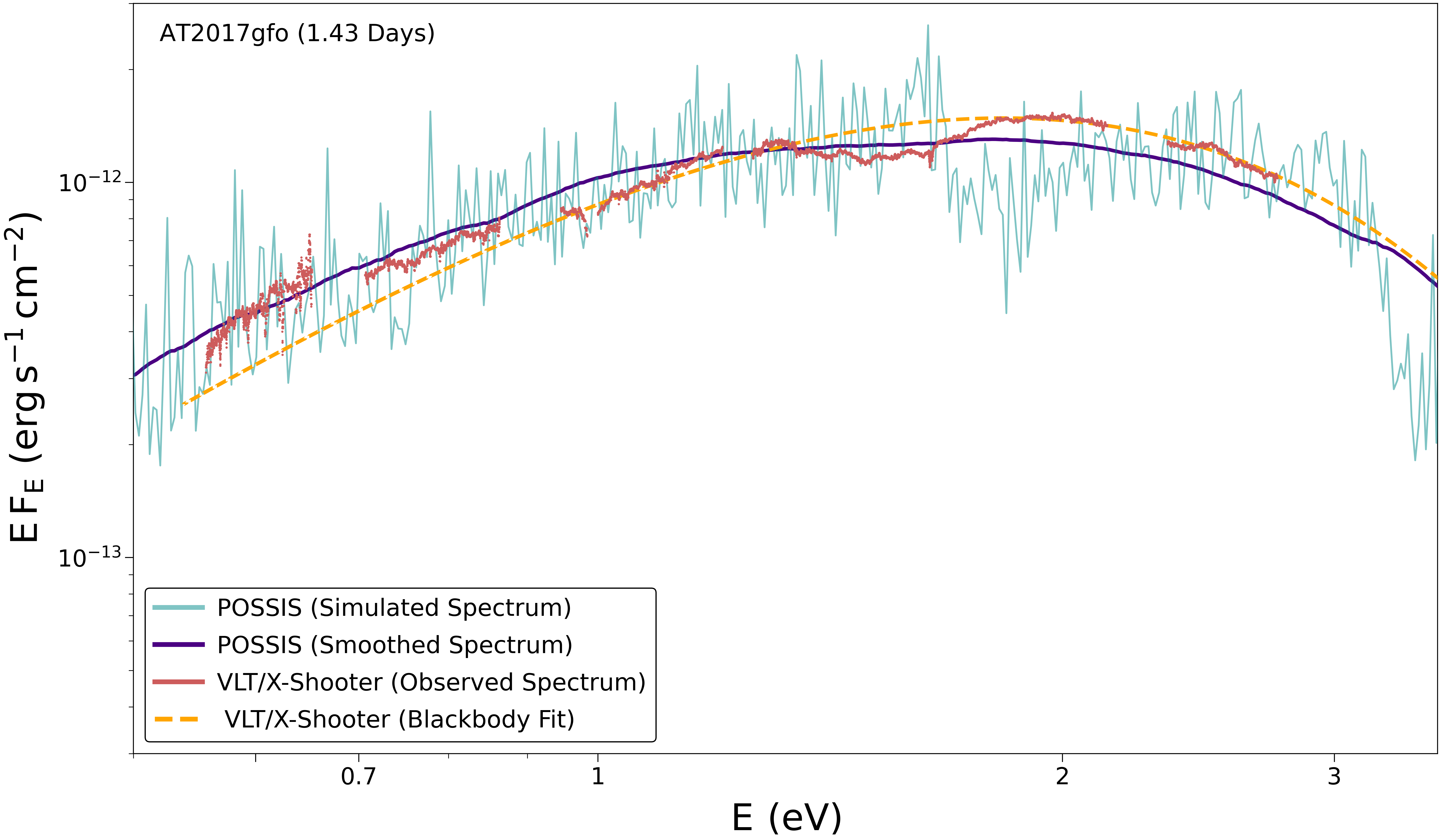}
\caption{Comparison between the observed \knname spectrum at epoch $\epoch =1.43$\,days and the synthetic POSSIS spectrum obtained with the \texttt{fit\_epliso} diagnostic.}
\label{nu_fnu_POSSIS}
\end{figure}

The isotropic luminosity $\liso$ and the peak energy $\ep$ are computed for each synthetic kilonova spectrum. 
Specifically, they are calculated from a smoothed version of the synthetic $E\, F_E$ spectra obtained using a third-order Savitzsky-Golay filter with a window of 250 pixels, with a pixel of $\sim35$\,\AA\, in the optical range. 
We note that, since we are solely concerned with the blackbody component of the observed spectrum, we can safely ignore spectral complexities arising from various effects, such as non-local thermal equilibrium effects \citep{Pognan2022Opcaity, Pognan2023NLTE} and fluorescence effects from individual elements in the kilonova ejecta \citep{Shingles2023Fluorescence}. 
Figure \ref{nu_fnu_POSSIS} shows a \texttt{POSSIS} spectrum at $1.43$\,days.

\begin{figure}[!h]
\centering
\includegraphics[width=0.96\textwidth]{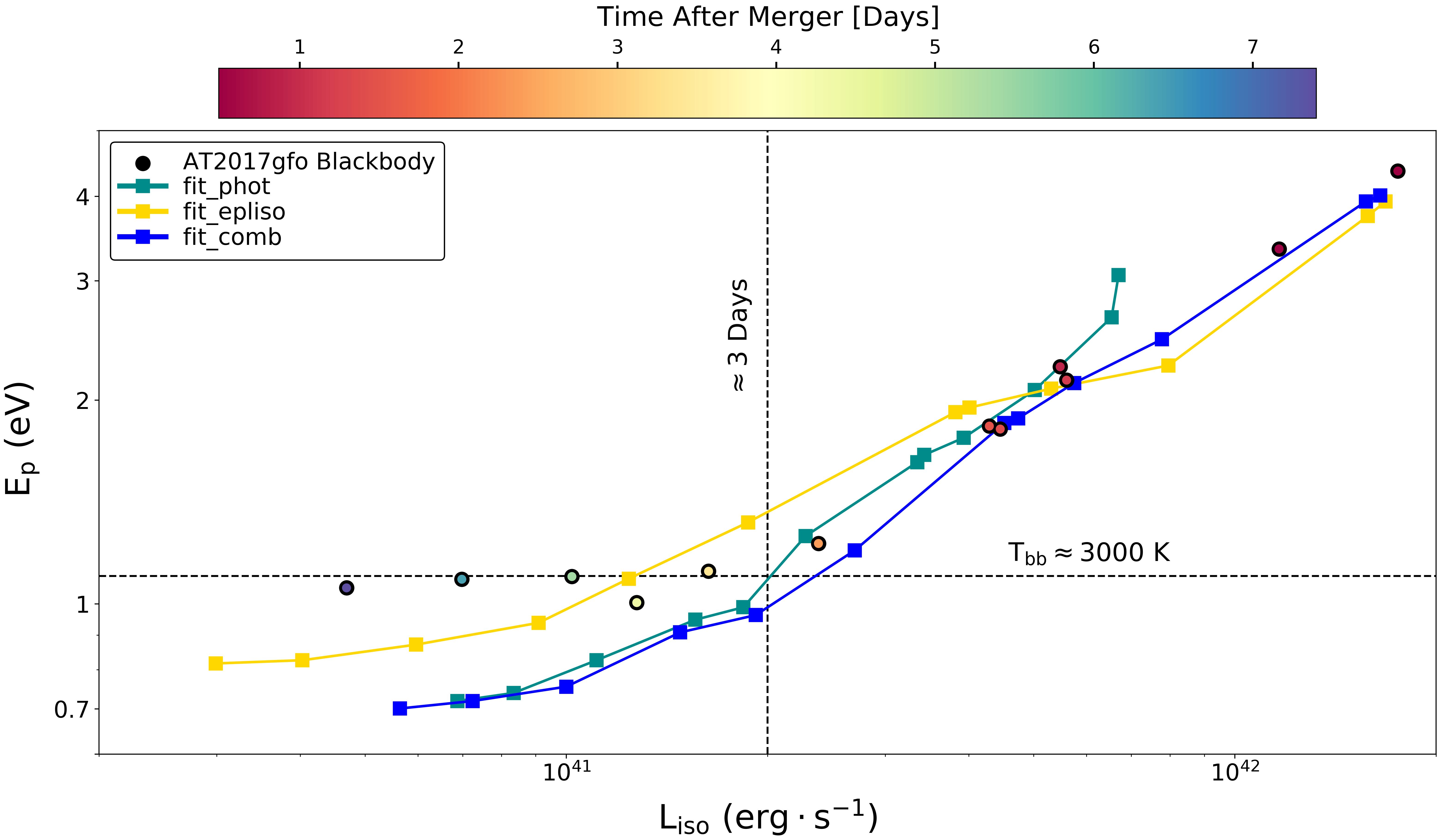}
\caption{Comparison between the $\ep-\liso$ correlation obtained from the blackbody component of the \knname spectra and the three different \texttt{POSSIS} best-fit models, namely \texttt{fit\_phot}, \texttt{fit\_epliso}, and \texttt{fit\_comb}. Table \ref{tab:bf_possis} reports the \texttt{POSSIS} best-fit parameters.}
\label{fig:knlc}
\end{figure}

Here, we focus on three particular \texttt{POSSIS} simulations, namely \texttt{fit\_phot}, \texttt{fit\_epliso} and \texttt{fit\_comb}. 
These correspond to the simulated spectra that best reproduce (i) the observed photometric light curves, (ii) the observed $\ep-\liso$ correlation, and (iii) the combination of these two observables, respectively.
By using a standard $\chi^2$ minimization process, the best-fit \knname model from \texttt{POSSIS} is selected with three different diagnostics: 
\begin{itemize}
    \item[1.] \texttt{fit\_phot}\ \ \ The synthetic photometric light curves are directly compared with the observed multi-band photometric data. $$\chi^2_{\texttt{phot}} = \sum_{i,j}\bigg(\frac{f_{\rm mod}^{i,j}-f_{\rm obs}^{i,j}}{\sigma_{\rm obs}^{i,j}}\bigg)^2$$
    where $f_{\rm obs}^{i,j}$ and $\sigma_{\rm obs}^{i,j}$ respectively represent the flux measurements and the related uncertainties at a given time $i$ and in a given filter $j$, while $f_{\rm mod}^{i,j}$ is the synthetic counterpart from \texttt{POSSIS}.\\
    \item[2.] \texttt{fit\_epliso}\ \ \ The $\ep$ and $\liso$ values derived from the synthetic spectra are fit to the observed values across the different observational epochs.
    $$\chi^2_{\texttt{epliso}} = \sum_{i}\bigg(\frac{E_{\rm p,mod}^{i}-E_{\rm p,obs}^{i}}{\sigma_{\rm Ep,obs}^{i}}\bigg)^2 +  \sum_{i}\bigg(\frac{L_{\rm iso,mod}^{i}-L_{\rm iso,obs}^{i}}{\sigma_{\rm Liso,obs}^{i}}\bigg)^2$$
    \item[3.]  \texttt{fit\_comb}\ \ \ By leveraging the complementary information contained in the photometric and spectroscopic data, the synthetic photometric light curves and the related $\ep$ and $\liso$ values are simultaneously compared to their observational counterpart.
    $$\chi^2_{\texttt{comb}} = \chi^2_{\texttt{phot}} + \chi^2_{\texttt{epliso}}$$
\end{itemize}
\noindent
Figure \ref{fig:knlc} compares the observed $\ep-\liso$ correlation against the predictions of the different \texttt{POSSIS} simulations, whose results are reported in Table \ref{tab:bf_possis}.

Simulated and observed $\ep-\liso$ correlations agree remarkably well, with early epochs exhibiting a similar power-law decay. Although it appears at later epochs and lower peak energies compared to the observations, \texttt{POSSIS} simulations predict a late-time flattening of the correlation, about $5\,$ days after the merger and at a peak energy around $0.7-0.8$\,eV.

\begin{table}[!h]
\caption{$\ep$ and $\liso$ best-fit values from the three \texttt{POSSIS} diagnostics. Units for the two quantities are eV and $10^{41}$ erg~s$^{-1}$, respectively.} 
\centering
\begin{tabular}{l l l l l l l}
\hline
\addlinespace[0.5em]
\multirow{2}{*}{} & \multicolumn{2}{l}{\texttt{fit\_phot}} & \multicolumn{2}{l}{\texttt{fit\_epliso}} & \multicolumn{2}{l}{\texttt{fit\_comb}} \\
\cmidrule(lr){2-3}  
\cmidrule(lr){4-5}  
\cmidrule(lr){6-7}
t (days) & $\liso$ & $\ep$ & $\liso$ & $\ep$ & $\liso$ & $\ep$ \\
\addlinespace[0.5em]
\hline
\addlinespace[0.5em]
0.488 & $6.70$ & 3.06 & $16.8$ & 3.93 & $16.5$ & 4.01 \\
0.517 & $6.54$ & 2.65 & $15.8$ & 3.74 & $15.7$ & 3.93 \\
0.919 & $5.02$ & 2.07 & $7.95$ & 2.25 & $7.78$ & 2.46 \\
1.158 & $3.93$ & 1.76 & $5.31$ & 2.08 & $5.75$ & 2.12 \\
1.458 & $3.43$ & 1.66 & $4.01$ & 1.95 & $4.74$ & 1.88 \\
1.458 & $3.35$ & 1.62 & $3.82$ & 1.92 & $4.52$ & 1.85 \\
2.448 & $2.28$ & 1.26 & $1.87$ & 1.32 & $2.70$  & 1.20 \\
3.459 & $1.84$ & 0.99 & $1.24$ & 1.09 & $1.92$ & 0.96 \\
4.356 & $1.56$ & 0.95 & $0.91$ & 0.94 & $1.48$ & 0.91 \\
5.485 & $1.11$ & 0.83 & $0.59$ & 0.87 & $1.00$ & 0.75 \\
6.520 & $0.83$ & 0.74 & $0.40$ & 0.83 & $0.72$ & 0.72 \\
7.316 & $0.69$ & 0.72 & $0.29$ & 0.82 & $0.56$ & 0.70 \\
\addlinespace[0.5em]
\hline
\end{tabular}
\label{tab:bf_possis}
\end{table}

From the comparison between \texttt{POSSIS} simulations and the \knname spectra, we identify two main arguments that corroborate the validity of the proposed correlation as a diagnostic tool and its interpretation in terms of kilonova physics.

\begin{figure}[!b]
\centering
\includegraphics[width=0.97\textwidth]{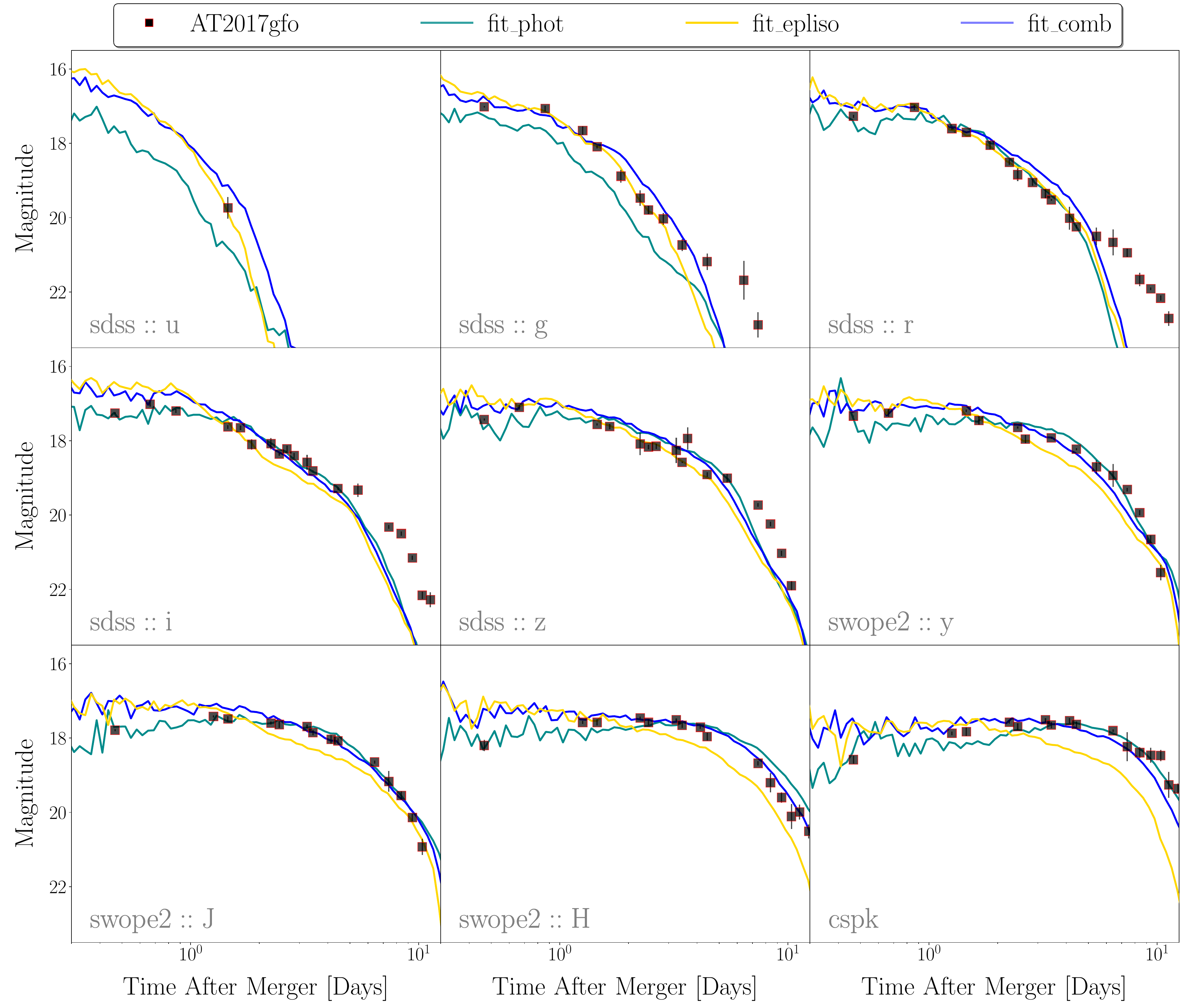}
\caption{Comparison between the \knname photometric observation and the \texttt{POSSIS} predictions obtained with the \texttt{fit\_phot}, \texttt{fit\_epliso} and \texttt{fit\_comb} diagnostics. For each of them, the corresponding ejecta parameters are provided in Table \ref{tab:POSSIS_BestFit_Theta25}.
Data considered in this analysis are the one compiled in \cite{Coughlin2018} and reported in \citep{Bulla2023POSSIS} with the full reference list.}
\label{fig:knlc1}
\end{figure}

First, although the \texttt{fit\_epliso} diagnostic is optimized to reproduce the observed $\ep-\liso$ correlation, the photometric light curves associated to the synthetic spectra closely match the observational data points in various photometric bands ($u,\,g,\,r,\,i,\,z,\,y,\,J,\,H,\,\text{and}\,K$ bands, see Figure \ref{fig:knlc1}). In particular, this diagnostic achieves an excellent match in the bluest bands (i.e., the $u$, $g$, and $r$ bands), where the correspondence between simulation and observational data points appears to be even better than that obtained with the \texttt{fit\_phot} method.
This result highlights the relevance of the \texttt{fit\_epliso} diagnostic that, by fitting only the peak energy $\ep$ and isotropic luminosity $\liso$ from the observed spectra, can accurately forecast the light curve evolution -- except for prominent spectral features that contaminate the spectra, especially at later epochs \citep{Waxman2018bb}.

Second, although (especially at early epochs) the synthetic spectra from the photometric light curve fit, namely \texttt{fit\_phot}, exhibit a substantial deviation to the observed correlation, this diagnostic still produces an $\ep-\liso$ correlation whose overall trend mirrors the observed one. 
After a rapid decline within 3--4\,days after the BNS merger, the correlation flattens around a $\ep$ constant value, marking the onset of the receding photosphere phase.
On the contrary, the \texttt{fit$\_$epliso} diagnostic provides a better fit to the observed $\ep-\liso$ correlation, with a remarkable match within the first three days after the BNS merger.
In this regard, the \texttt{fit\_comb} diagnostic provides a trade-off between the two previous approaches. Notably, the \texttt{fit\_comb} prediction of the $\ep-\liso$ correlation tightly follows the \texttt{fit\_epliso} one within the first day after the BNS merger, while at $\epoch > 3$ days the \texttt{fit\_phot} diagnostic dominates.

\begin{table*}[!t]
\captionsetup{width=\textwidth}
\caption{\texttt{POSSIS} best-fit parameters for \knname, at a fixed viewing angle of $\thetaobs= 25^{\circ}$.}
\centering
\begin{tabular}{llll}
\hline
\addlinespace[0.5em]
Parameters & \texttt{fit\_phot} & \texttt{fit\_epliso} & \texttt{fit\_comb}  \\ 
\addlinespace[0.5em]
\hline
\addlinespace[0.5em]
$m_{\rm dyn}\,[M_\odot]$ & 0.02 & 0.01 & 0.02    \\ 
$\bar{v}_{\rm dyn}\,[c]$ & 0.12 & 0.15 & 0.15  \\ 
$\bar{Y}_{e,\,\rm dyn}$ & 0.20 & 0.30 & 0.25   \\ 
$m_{\rm wind}\,[M_\odot]$ & 0.13 & 0.09 & 0.09     \\ 
$\bar{v}_{\rm wind}\,[c]$ & 0.05 & 0.03 & 0.03  \\ 
$\bar{Y}_{e,\,\rm wind}$ & 0.40 & 0.40 & 0.40  \\
\addlinespace[0.5em]
\hline
\end{tabular}
\label{tab:POSSIS_BestFit_Theta25}
\end{table*}

The \texttt{POSSIS} best-fit parameters of the \knname ejecta retrieved from the three different diagnostic tools are summarized in Table \ref{tab:POSSIS_BestFit_Theta25}.
We note that, compared to the standard \texttt{fit\_phot} case, the \texttt{fit\_epliso} diagnostic favors a less massive dynamical ejecta that expands more rapidly and has a higher electron fraction. Also, \texttt{fit\_epliso} yields a lighter wind ejecta component that expands more slowly but retains a similar electron fraction.
From the analysis of the \texttt{fit\_comb} results, we can determine which diagnostic each \texttt{POSSIS} parameter is most sensitive to.
While the \texttt{fit\_epliso} diagnostic determines the $\bar{v}_{\rm dyn}$, $\bar{v}_{\rm wind}$, and $m_{\rm wind}$ parameters, the $m_{\rm dyn}$ best-fit value is set by the \texttt{fit\_phot} diagnostic. 
Also, the $\bar{Y}_{e,\,\rm dyn}$ value results to be the average of the \texttt{fit\_epliso} and \texttt{fit\_phot} best-fit values, while $\bar{Y}_{e,\,\rm wind}$ remains constant across all diagnostics.
In summary, the \texttt{fit\_epliso} diagnostic primarily determines the wind parameters, while the dynamical parameters emerge from the combination of both diagnostics.
Clearly, the primary limitation in estimating kilonova ejecta properties stems from the sampling over a fixed-step grid, as demonstrated by the saturation of the $m_{\rm dyn}$ and $\bar{Y}_{e,\,\rm wind}$ parameters over the boundaries of the sampling grid, i.e., 0.02\,$c$ and 0.40, respectively.
To obtain more accurate best-fit values and meaningful uncertainties, one would need to expand the grid and implement an interpolation scheme that can generate spectra for any combination of the kilonova ejecta parameters.
Although this approach demands substantial computational resources, it highlights the need for further investigation of the $\ep-\liso$ correlation as a diagnostic tool for kilonova transients, which promises to provide fundamental insights into the physics of these extraordinary phenomena.

\section{Outlooks and Conclusions}

It is worth noting that assessing the systematics affecting the $\ep-\liso$ correlation inevitably requires more kilonova events and a detailed study of the effects of each prior assumption. 
In this respect, we discuss the robustness of our approach through two primary arguments. 

First, the accuracy of the $\ep-\liso$ correlation is intrinsically linked to the goodness of the spectral fitting over the observed spectra. 
As shown in Figure \ref{nuFnu_Fit_Residual}, our spectral modeling consistently provides a reasonable match to the observed \knname spectra, thus guaranteeing an accurate estimation of the energy peak $\ep$ and the isotropic luminosity $\liso$ of the kilonova.

Second, similar conclusions on the $\ep-\liso$ correlation of \knname can be obtained from the literature reference that independently analyzed this kilonova event \citep[][see Table B2 and B3]{Waxman2018bb}. Specifically, since the bolometric luminosity of the observed blackbody and its effective temperature $T_{\text{eff}}$ exhibit a strong temporal correlation, extracting the $\ep-\liso$ correlation is rather straightforward. In practice, by applying Wien’s displacement law to $T_{\text{eff}}$, one can determine the peak wavelength of the blackbody spectrum at each observational epoch and, from that, derive the corresponding photon energy in eV. This value serves as a reliable proxy for the peak energy $\ep$. However, the limitation is that it does not account for the Doppler factor or redshift corrections (which are instead considered in the \texttt{grbjet} model). Nevertheless, it is evident that, despite the two different methods applied to \knname, both analyzes consistently reveal the $\ep-\liso$ correlation, which is characterized by a rapid decay within the first three days after the BNS merger, and a subsequent flattening at an almost constant value of $\ep$.

More intriguingly, photometric observations of other kilonova transients \citep{Troja22, Yang2024} have revealed a temporal correlation of the bolometric luminosity and the effective blackbody temperature, similar to the \knname case. These results ultimately suggest that different kilonovae may trace a common trajectory in the $\ep-\liso$ plane, which -- if confirmed -- would represent a remarkable achievement in our understanding of such transient events. However, it is crucial to emphasize that, without additional spectroscopic follow-up, extracting the actual $\ep-\liso$ correlation becomes difficult as the photometric observations can be strongly contaminated by any emission mechanisms that deviate from a pure blackbody.  

\begin{figure}
\centering
\includegraphics[width=\textwidth]{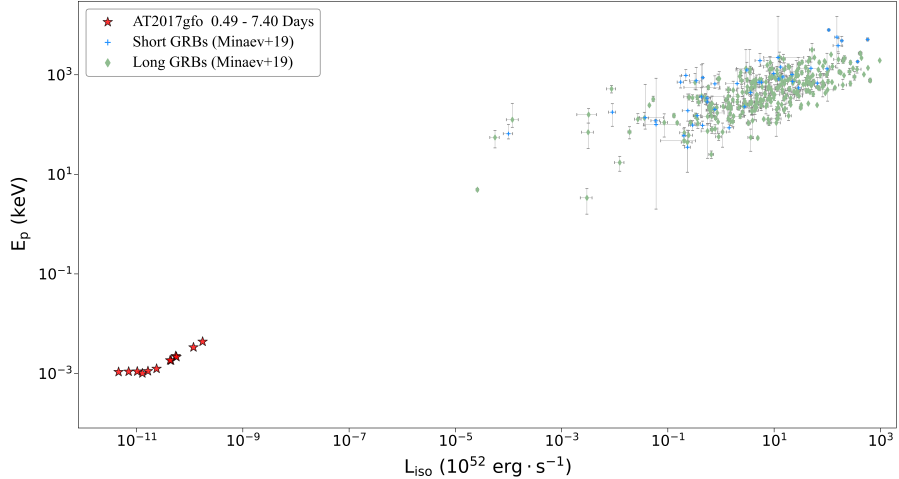}
\caption{The observed $\ep-\liso$ from \knname kilonova transient is compared to the measurements obtained for short and long gamma-ray bursts (adapted from \cite{Minaev2020}), showing an apparent alignment with the Short and Long GRBs.}
\label{kn_vs_literature}
\end{figure}

From a more general perspective, we note that empirical correlations represent a powerful tool for studying the radiative properties of another kind of transient events, namely long and short GRBs \citep{amati2002, yonetoku2004, ghirlanda2004, amati2006, frontera2013, Zhang19, maistrello24_aa}.
Indeed, the same $\ep-\liso$ correlation exists for the prompt emission of GRBs in the high-energy (from keV to MeV) and high-luminosity ($>10^{48}$ erg~s$^{-1}$) domain. 
In Figure \ref{kn_vs_literature}, we observe that \knname populates the low-energy extension of the $\ep-\liso$ correlation observed among the long and short GRBs\,\footnote{$\liso$ should not be confused with the isotropic peak luminosity used more often in GRB studies, for example in the Yonetoku relation.}.
For adapting the literature results \citep{Minaev2020} to our work, we derive the $\liso$ values of both long and short GRBs by relying on the estimation of the isotropic energy $E_{\text{iso}}\,$ and the time interval $T_{90}$ over which 90\% of the total GRB emission is observed.
This unanticipated alignment may hint at a common underlying radiative process among these distinct transients. In fact, several studies have identified potential evidence for a photospheric component in the prompt emission of GRBs \citep{Meszaros00, Zhang02, Ryde06, Giannios06, Thompson2007, Ryde10, Zhang11, Ito19}, thereby motivating further investigation into whether a unified framework exists in which the $\ep-\liso$ correlation serves as a bridge between the physics of kilonova transients and gamma-ray burst events.

In conclusion, starting from a fresh perspective rooted in X/$\gamma$-ray astronomy and applying it to the time-resolved spectroscopic observations of \knname, our new analysis of the temporal evolution of the spectral properties of the \knname kilonova transient has revealed an underlying correlation between two key observables: the peak energy of the $E\,F_E\,$ spectrum and the isotropic-equivalent luminosity. The robust agreement between observations, spectral modeling, and radiative-transfer simulations indicates that the $\ep-\liso$ correlation is a promising diagnostic tool for probing the properties of kilonova ejecta and potentially would significantly enhance our understanding of the post-merger evolution in BNS events (see,  e.g.,~\cite{Baiotti2016} for a review). 

Looking ahead, a refined application of the $\ep-\liso$ diagnostic across a broader sample of GRB-associated kilonova events will be presented in a forthcoming study, where kilonova physical parameters inferred from light curve observations will be used to construct simulated spectra, leading to the computation of the $\ep-\liso$ correlation and exploration of its true impact on such transients.

\newpage

\bibliographystyle{elsarticle-harv}
\bibliography{bibliography}

\end{document}